\documentclass[pra,10pt,twocolumn,showpacs]{revtex4}%
\usepackage{amsmath}
\usepackage{amsfonts}
\usepackage{amssymb}
\usepackage{graphicx}%
\begin{document}
\title{Implement Quantum Random Walks with Linear Optics Elements}
\author{Zhi Zhao}
\author{Jiangfeng Du}
\author{Hui Li}
\author{Tao Yang}
\author{Zeng-Bing Chen}
\author{Jian-Wei Pan}
\affiliation{Department of Modern Physics, University of Science and Technology of China,
Hefei, 230027, People's Republic of China}

\begin{abstract}
The quantum random walk has drawn special interests because its remarkable
features to the classical counterpart could lead to new quantum algorithms. In
this paper, we propose a feasible scheme to implement quantum random walks on
a line using only linear optics elements. With current single-photon
interference technology, the steps that could be experimentally implemented
can be extended to very large numbers. We also show that, by decohering the
quantum states, our scheme for quantum random walk tends to be classical.

\end{abstract}
\maketitle

That quantum physics differs from classical physics is due to the quantum
coherence, which has been utilized to perform quantum computation such as
Shor's factoring algorithm \cite{shor} and Grover's database search algorithm
\cite{grover}. However finding quantum algorithms is very difficult. Recently,
several groups have investigated the quantum random walk, with the hope that
it might help construct new quantum algorithms \cite{aha,far,Na,7,8,Tra,10,du}%
. Indeed, the first quantum algorithms based on quantum walks with an
exponential speedup have been reported \cite{al}. Further, Travaglione
\textit{et al.} proposed a scheme to implement a discrete-time quantum random
walk by an ion trap quantum computer\textit{\ }\cite{Tra} and D\"{u}r
\textit{et al.} proposed to use neutral atoms trapped in optical lattices to
realized such quantum walks. The first experimental implementation of the
quantum random walk algorithm was reported by Du \textit{et al.} \cite{du}.

In this paper, we propose a feasible scheme to implement quantum random walks
on a line using only linear optics elements. With the defined movement and
\textit{the dynamic line}, we demonstrate that our scheme fulfills all
requirements of the quantum random walks. With current single-photon source
and its interference technology, our scheme is at the reach of current
experiment. The remarkable difference between the quantum and classical random
walks could be investigated with a larger number of steps. Also, by decohering
the quantum states, our scheme for quantum random walk tends to be classical.

Let us consider a particle performing the discrete-time quantum random walks
on a line. Besides the position degrees of freedom, the particle has an
additional degree of freedom, namely the \textquotedblleft
chirality\textquotedblright, which takes value \textquotedblleft
Left\textquotedblright\ and \textquotedblleft Right\textquotedblright. The
walk by such a particle can be described as follows: at every time step, its
chirality undergoes the Hadamard transformation and then the particle moves
according to its (new) chirality state. In details, the chirality state is
denoted by a vector in the Hilbert space spanned by $\left\vert \text{L}%
\right\rangle $ (Left) and $\left\vert \text{R}\right\rangle $ (Right). If the
chirality state is $\left\vert \text{L}\right\rangle $ the particle moves to
the left, while if the chirality state is $\left\vert \text{R}\right\rangle $
the particle moves to the right. Generally, the chirality state might be a
superposition of $\left\vert \text{L}\right\rangle $ and $\left\vert
\text{R}\right\rangle $, this leads the particle moving coherently to both the
left and the right. It is this quantum coherence that enables the quantum
random walk to remarkably outperform the classical walk.

In order to implement this quantum random walk, the essential idea is to
implement the movement depending upon its \textquotedblleft
chirality\textquotedblright\ and to perform the Hadamard operation for
obtaining further \textquotedblleft chirality\textquotedblright\ for the next
step. In our scheme we use the polarization states of photons to represent the
\textquotedblleft chirality\textquotedblright. We represent the
\textquotedblleft chirality\textquotedblright\ state $\left\vert
\text{R}\right\rangle $ by the horizontal polarization photon state
$\left\vert \text{H}\right\rangle $, and $\left\vert \text{L}\right\rangle $
by vertical polarization state $\left\vert \text{V}\right\rangle $. The
implementation is based on simple linear optical elements, \textit{e.g.}
polarization beam splitters (PBS) and half-wave plates (HWP). When a
superposition state passes through a PBS as shown in FIG. 1a, the PBS will
direct the photon into two output ports because it transmits only horizontal
component and reflects only vertical component. In order to construct the
equivalence to the quantum random walk on a line, we could denote the
transmission as \textquotedblleft Right\textquotedblright\ side and the
reflection as the \textquotedblleft Left\textquotedblright\ side when photons
directly pass through the PBS. However,when a superposition state directly
passes through a PBS from the \textquotedblleft Left\textquotedblright\ side,
it will meet some difficulty since the horizontal component is transmitted and
the vertical component is reflected. Thus we need a \textquotedblleft
modified\textquotedblright\ PBS (the $\overline{\text{PBS}}$) which transmits
the vertical component and reflects the horizontal component. The
$\overline{\text{PBS}}$ could be realized by rotating the polarization of the
photon by $90^{0}$ with a HWP (let the angle between the HWP axis and
horizontal be $45^{0}$) followed by rotating back with two HWP's after passing
through the PBS (R$_{90}$ as shown in FIG. 1b). After passing through the
$\overline{\text{PBS}}$, the horizontal component moves to the
\textquotedblleft Right\textquotedblright\ and the vertical component to the
\textquotedblleft Left\textquotedblright. Also, because there are different
components from the two directions, one could conveniently superpose them by
one PBS as shown in FIG. 1c. Therefore, with the PBS and the $\overline
{\text{PBS}}$ (the modified PBS), we defined the photonic movement depending
upon its \textquotedblleft chirality\textquotedblright, \textit{i.e.}
polarization degree of freedom of the photons.%

\begin{figure}
[ptb]
\begin{center}
\includegraphics[
height=6.2714cm,
width=6.9172cm
]%
{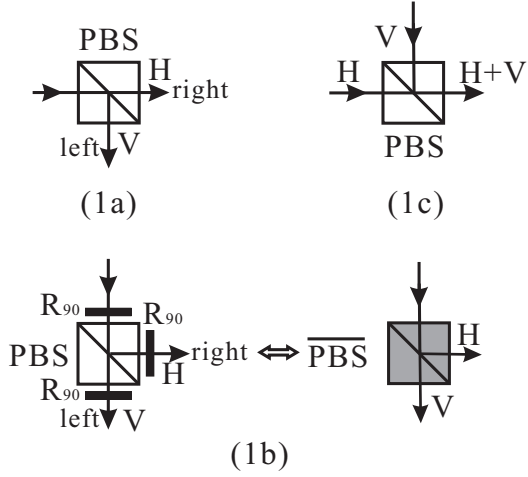}%
\caption{Schematic of the setup to make use of polarization beam splitter
(PBS) and half-wave plate (HWP) as the basic elements for implementing the
quantum random walk on a line.}%
\label{Fig1}%
\end{center}
\end{figure}

During the quantum random walk, one needs to perform a Hadamard operation in
each step. This can easily be implemented by rotating the photonic
polarization by $45^{0}$ with a HWP (by letting the angle between the HWP axis
and horizontal be $22.5^{0}$). After passing the HWP (R$_{45}$) it will
undergo the following Hadamard transformation,%

\begin{equation}
\left\vert \text{H}\right\rangle \rightarrow\frac{1}{\sqrt{2}}\left(
\left\vert \text{H}\right\rangle +\left\vert \text{V}\right\rangle \right)  ,
\end{equation}

\begin{equation}
\left\vert \text{V}\right\rangle \rightarrow\frac{1}{\sqrt{2}}\left(
\left\vert \text{H}\right\rangle -\left\vert \text{V}\right\rangle \right)  .
\end{equation}
%

\begin{figure}
[ptbh]
\begin{center}
\includegraphics[
height=12.591cm,
width=8.6108cm
]%
{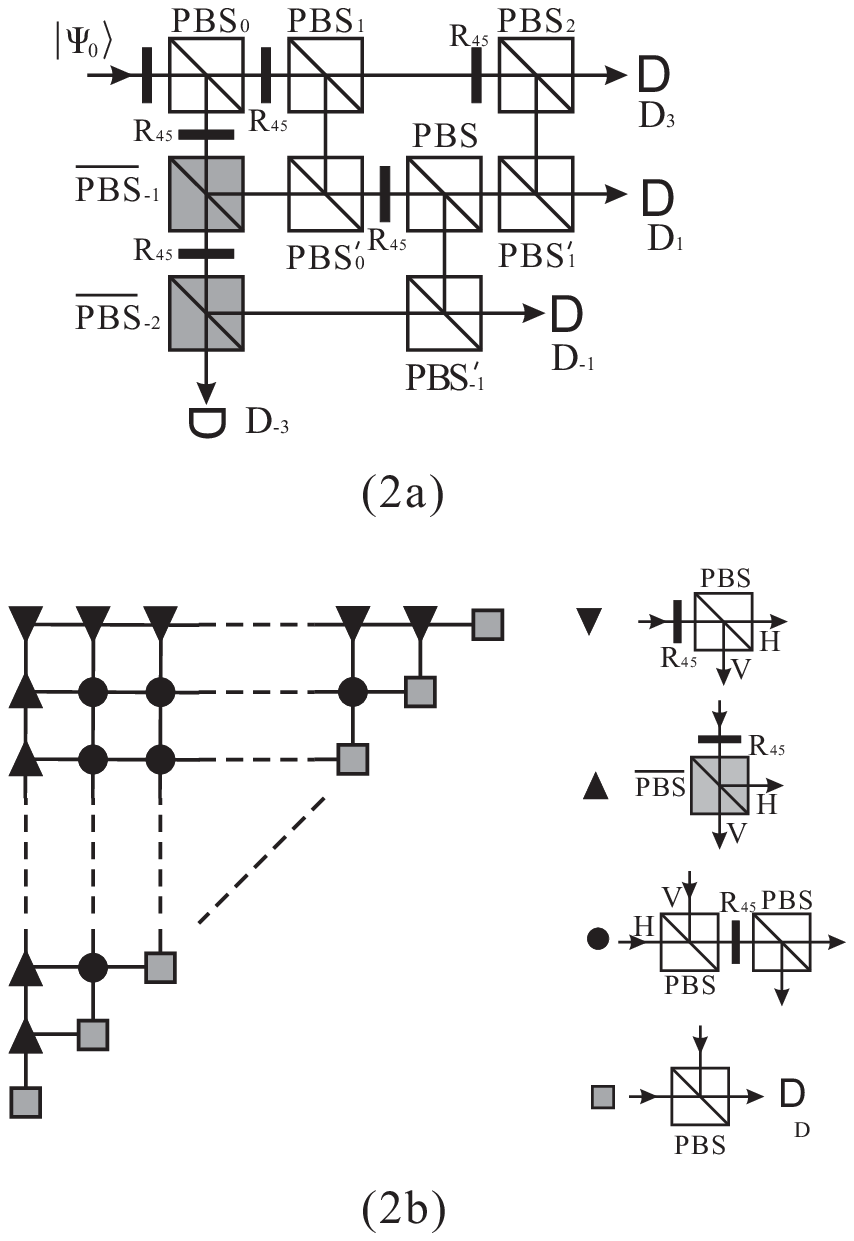}%
\caption{Schematic of the setup to implement quantum random walk on a line
with three steps (2a) and the optical network with $N$ steps (2b).}%
\end{center}
\end{figure}

With the defined movement and the Hadamard operation, we can construct the
optical network to implement the quantum random walk on a line. Consider the
arrangement shown in FIG. 2a, each PBS or $\overline{\text{PBS}}$ denotes an
integer point on the line. After the initial polarization photon state
$\left\vert \Psi_{0}\right\rangle $ undergoes a Hadamard transformation and
then passes through the PBS$_{0}$, the horizontal component moves to the right
and the vertical component to the left. Then each $45^{0}$ rotation of the
polarization for two components produces the new chirality so that it could
proceed the next walk. The transmitted components move to the integer $2$ and
$-2$ respectively, and the horizontal component reflected from $\overline
{\text{PBS}}_{-1}$ and the vertical component from PBS$_{1}$ will overlap at
the PBS$_{0}^{^{\prime}}$. Here it should be noted that, in order to let all
components move forward and not mix with further input states, we shall
introduce a new PBS$_{0}^{^{\prime}}$ to replace the PBS$_{0}$, which is
equivalent to move the integer $0$ to $0^{\prime}$. The integers $1$,
$0^{\prime}$, $-1$ compose a new line, leading to convenient measurement after
the random walk, as will be explained later. The state coming out from
PBS$_{1}$, PBS$_{0}^{^{\prime}}$ and $\overline{\text{PBS}}_{-1}$ further
undergoes a Hadamard operation and then performs the next step of the walk.
The transmitting components from PBS$_{2}$ and $\overline{\text{PBS}}_{-2}$
will move to the position $3$ and $-3$ respectively. The two components
reflected from PBS$_{2}$ and PBS$_{0}^{\prime}$ will overlap at PBS$_{1}%
^{\prime}$ and the other components reflected from $\overline{\text{PBS}}%
_{-2}$ and PBS$_{0}^{\prime}$ will overlap at PBS$_{-1}^{\prime} $.
Introducing PBS$_{1}^{\prime}$ and PBS$_{-1}^{\prime}$ to replace the
PBS$_{1}$ and $\overline{\text{PBS}}_{-1}$ has the same reason as introducing
the PBS$_{0}^{\prime}$. The integers $3$, $1^{\prime}$, $-1$, and $-3$
assemble a new line and one can position four detectors behind them to measure
the probability distribution after three steps. By iterating the above steps,
we could implement random walk on a line with the unlimited steps in principle
by simply adding more PBS's, $\overline{\text{PBS}}$'s and HWP's. This optical
network for quantum random walk for arbitrary $N$ steps is shown in Fig. 2b.

The essential idea in our scheme is to introduce \textit{the dynamic line},
which is equivalent to the original line on which the walk is performed. With
the moving integer positions we could fulfill all the requirement for the
quantum random walk in a line and the scheme provide us a convenient method to
measure the distribution after certain steps. Furthermore, it is easily found
that the general quantum random walk proposed in Ref. \cite{Na} can also be
implemented in our scheme, by rotating the photonic polarization by arbitrary
angle $\theta$ with the HWP (R$_{\theta}$) rather than $45^{0} $ when
performing the Hadamard transformation. Note that in this case the angle
between the HWP axis and horizontal should be $\theta/2$.

So far, we have described the quantum random walk on a line. We now discuss
how to realize it experimentally. To implement the quantum random walk, the
source that produces single photons is demanded. It is available with
quantum-dot single photon sources \cite{kim,kur,lou,fod} or with parametric
down conversion \cite{kwiat} by performing a measurement on one photon. In the
quantum random walk, the probability distribution is strongly dependent upon
its initial state. The general input $\left\vert \Psi_{0}\right\rangle
=\cos\theta\left\vert \text{H}\right\rangle +e^{i\varphi}\sin\theta\left\vert
\text{V}\right\rangle $ can be easily obtained by letting single-photon
sources pass through two QWP's and a HWP \cite{englert}. The key requirement
for the experimental realization is to overlap the horizontal component and
the vertical one on the PBS in order to utilize the quantum coherence. This
could be achieved, such as in the Rome teleportation experiments
\cite{martini}. Finally, the features of the quantum walks could be
investigated by positioning detectors behind the PBS's at the final step. All
these requirements have been in the reach of the current technology of linear
optics. Also, since the single photon source is very bright, the steps for the
quantum random walk are scalable. This is comparative to other scheme such as
ion trap \cite{Tra} or NMR system \cite{du} with only a few steps. Therefore,
with our scheme one can investigate quantum random walks with any initial
state for a large number of steps. It should be noted that, whenever taking
odd steps or even steps, we need only measure the distribution in the odd
positions or in even positions, respectively. But it will not lead to any
contradiction because it is obvious that, for either the quantum random walk
or the classical one, the probability is zero at even (odd) positions when
taking odd (even) steps.

As described by many authors, the remarkably distinguished features between
the quantum random walk and the classical one stem from the quantum coherence.
However, one could implement the classical random walk by decohering quantum
states in our scheme. With slight modification of our scheme, this can be
realized by appropriately placing phase shifts more than coherence length of
single photon in the optical network, so that the state of the quantum walk
decoheres after each step. Therefore, our scheme could be fully used to
compare the remarkable features such as the probability distribution and the
standard deviation to the classical one for random walk on a line. Further, by
changing the phase shift between zero and the coherence length, it is also
possible to investigate the decoherence effect in the quantum random walk in experiments.

In conclusion, we have proposed a feasible scheme to implement the quantum
random walk on a line with linear optics elements. In the scheme we construct
\textit{the dynamic line} so that we could conveniently measure the
probability distribution after certain steps. With nowadays single photon
interference technology and highly precise linear optics elements, our scheme
are flexible and scalable, and could be implemented experimentally for any
initial state with a large number of steps. The experimental realization of
quantum random walk also provides us a valuable estimation for quantum
computer with linear optics.

This work was supported by the National Natural Science Foundation of China,
Chinese Academy of Sciences and the National Fundamental Research Program
(Grants No. 2001CB309303 and 2001CB309300).

\end{document}